# Topological luminophor $Y_2O_3$:$Eu^{3+}$+Ag with high electroluminescence performance

**Mingzhong Wang, Longxuan Xu, Guowei Chen, Xiaopeng Zhao***

*Smart Materials Laboratory, Department of Applied Physics, Northwestern Polytechnical University, Xi'an 710129, P.R.China.*

*\* xpzhao@nwpu.edu.cn*

**Abstract:** Improving luminescent intensity is a significant technical requirement and scientific problem for the luminescent performance of fluorophor materials through the ages. The process control and luminescence performance still limit the developments of luminescent intensity even through it can be improved partly by covering or magnetron sputtering of precious metals on the surface of the fluorophore materials. On the basis of the improvement of luminescence center radiative transition rate by surface plasma resonance and $Y_2O_3$:$Eu^{3+}$ microsheet phosphors, a fundamental model for topological luminophor $Y_2O_3$:$Eu^{3+}$+Ag was designed referencing the concepts of topological materials in order to enhance luminescent performance by composite-luminescence, which composed of $Eu^{3+}$centric electroluminescence and surface plasma-enhanced photoluminescence by Ag. The topological luminophor $Y_2O_3$:$Eu^{3+}$+Ag was successfully synthesized with an asymmetric-discrete Ag nanocrystal topological structure on the surface just via illumination. Experiment results suggest that the luminescence performance of topological luminophor $Y_2O_3$:$Eu^{3+}$+Ag increased by about 300% compared with that of $Y_2O_3$: $Eu^{3+}$ phosphors on the same conditions. The design of a topological luminophor provides a new approach to further improve the luminescent intensity of phosphors.

## 1 Introduction

Luminescent materials present promising application foreground in optoelectronic devices, panel display, biomedical analysis/imaging[1-3], quantum information technology[4] and improving superconducting properties[5-7]. Therefore, enhancing the luminescent intensity is a hot spot for research [8-10]. Given the abundant f-orbital configurations, rare earth (RE) luminescent materials have received considerable research interest among all types of luminescent materials which is well-knowed for its favourable stability and luminescence dating [11,12]. To further improve the performance of RE luminescent materials, precious metal nanomaterials have attracted considerable attention for the possible coupling effect between surface plasma resonance and luminescent materials[13-16]. For example, Zhang et al.[17]

described a structure that sets SiO$_2$ as a spacer layer in order to enhance the luminescence performance of CeF$_3$: Tb$^{3+}$ and LaF$_3$: Yb$^{3+}$/Er$^{3+}$ based on surface plasma resonance of AuNRs. Song et al.[18] coated NaYF$_4$: Tb$^{3+}$ onto the nano Au or Ag to form a core-shell structure. The quenching efficiency of Au or Ag within the samples is decided by energy transfer efficiency that depends largely on the extent of the spectral overlap with the luminescence resonance energy transfer (LRET) system. Yu et al.[19] synthesized Y$_2$O$_3$:Eu$^{3+}$ nanotubes by hydrothermal method, with Pt nanoparticles onto the surface via vacuum evaporation, and the effciency of solar cells improved by approximately 11.96% with the increased luminescence intensity[20-23]. Precious metals are usually used as a core or shell to form core-shell structure which is usually applied to enhance the photoluminescence of RE luminescent materials. The theories of surface plasma resonance to improve electroluminescence performance[24-27] are currently used for organic light-emitting diode (OLED) devices[28-31]. The application of surface plasma resonance for improved luminescence performance under the electrice field of RE luminescent materials is rarely mentioned.

The last few decades have witnessed the topological insulator and topological superconductor as a frontier topic in the field of materials[32-37]. Zhang et al.[38] discovered the quantum Hall effect which existed in 2D topological insulators. Feasibility was verified by Chen et al.[39], who reported that Bi$_2$Te$_3$ can be treated as a 3D topological insulator. Sato and Ando[40] discussed the theories and developments of topological insulator and topological superconductor since the topological quantum system was introduced for the first time in 1982. However, the combination of topology and materials is one of the vital aspects in the field of materials in the years to come[41-44].

Herein, on the basis of improving the radiative transition rate of luminescence centers by surface plasma resonance, a fundamental model for topological luminophor Y$_2$O$_3$:Eu$^{3+}$+Ag is designed referencing the concepts of topological materials based on Y$_2$O$_3$:Eu$^{3+}$ microsheet phosphors. Notably, the topological luminophor Y$_2$O$_3$:Eu$^{3+}$+Ag was successfully synthesized with an asymmetric-discrete Ag nanocrystal topological structure on the surface just via illumination. In addition, the connections among the luminescent intensity of topological luminophors Y$_2$O$_3$:Eu$^{3+}$+Ag, Ag content and electric field was investigated and the mechanism of enhanced luminescence performance was also analyzed.

## 2 Experimental section

## 2.1 Topological luminophor model

Based on the hydrothermal preparation of Y$_2$O$_3$:Eu$^{3+}$ microsheet phosphors[45-47], a fundamental model for Y$_2$O$_3$:Eu$^{3+}$+Ag was designed referencing the concepts of topological materials which was prepared just via illumination (Fig. 1).

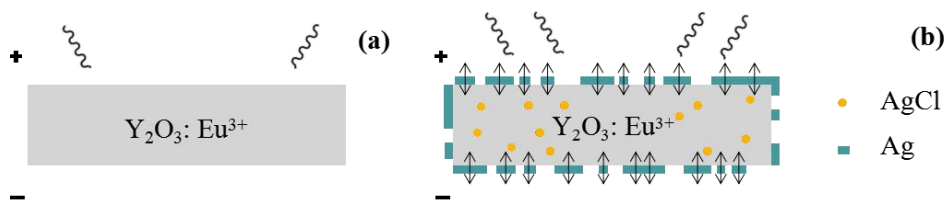

**Fig.1** Schematic model of the enhanced electroluminescence of topological luminophor $Y_2O_3$:$Eu^{3+}$+Ag: (a) $Y_2O_3$: $Eu^{3+}$ phosphor was excited to produce electroluminescence under an electric field and $Eu^{3+}$ acts as a luminescence center. (b) Based on $Y_2O_3$: $Eu^{3+}$ phosphor and the Ag nanocrystal topological structure onto the surface, $Eu^{3+}$ was treated as the luminescence center to produce electroluminescence under electric field including Ag nanocrystal topological structure to enhance photoluminescence of the samples to comply with surface plasma resonance. As a consequence, the luminescent performance of topological luminophor $Y_2O_3$:$Eu^{3+}$+Ag greatly improved.

## 2.2 Reagents

RE oxides ($RE_2O_3$, RE=Y and Eu, AR) and NaOH were supplied by Sinopharm Chemical Reagent Co., Ltd. Hydrochloric acid (AR, 36.5 wt%) and dehydrated ethanol (AR) were purchased from Tianjin Zhiyuan Reagent Company. AO (AR, 99%) was supplied by Tianjin Fuchen Chemical Industry Co., Ltd. and silver nitrate (AR) was purchased from Shanghai Institute of Fine Chemical Materials. No further purification was conducted and all chemicals were used as received.

## 2.3 Preparation

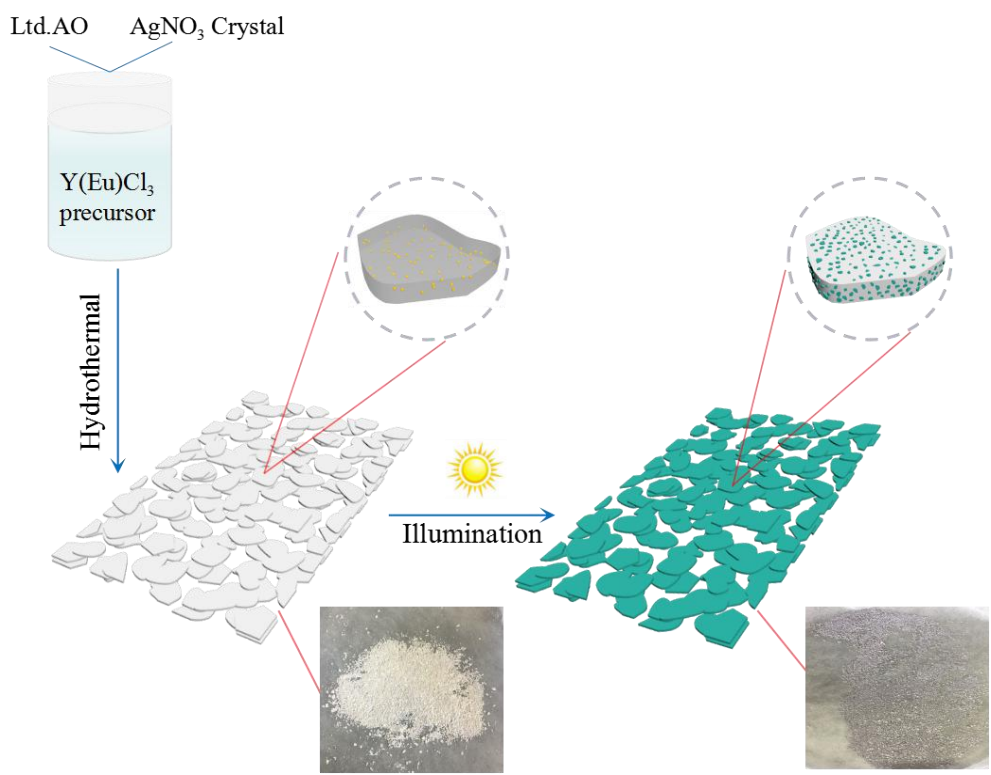

**Fig.2** Schematic representation of the preparation for topological luminophor $Y_2O_3$:$Eu^{3+}$+Ag. Note

that AgCl was mixed into $Y_2O_3$:$Eu^{3+}$ microsheets after hydrothermal reaction. The color of the samples changed from white to gray at macrolevel as AgCl decomposed into Ag and moved to the surface by illumination.

The topological luminophor $Y_2O_3$:$Eu^{3+}$+Ag was synthesized through illumination after using modified methods based on hydrothermal conditions (Fig. 2). In a typical procedure, 0.153g of $Y_2O_3$ powders and 0.025g of $Eu_2O_3$ powders (molar ratio = 0.95:0.05) were stirred and heated with hydrochloric acid. The milky-white solution was transformed into a viscous transparent solution after some time. Excess HCl was driven off by heating until the white crystals separated. The prepared Y(Eu)$Cl_3$ was dissolved into ammonium oxalate solution under constant magnetic stirring at 2 °C., $AgNO_3$ (molar ratios = 0.99: 0.01, 0.98: 0.02, 0.97:0.03, 0.96:0.04, 0.95:0.05, 0.94:0.06, 0.92:0.08, 0.90:0.10, 0.85:0.15) was added into the mixed solution with stirring for 30 min and pH was adjusted to 9. The mixture was transferred into a Teflon-lined autoclave for 2 h of solvothermal treatment at 160 °C. The autoclave was cooled to room temperature naturally, and the precipitates were centrifuged and washed several times, white $Y_2O_3$:$Eu^{3+}$+AgCl phosphors were prepared by calcination at 800 °C for 2 h after all. Finally, the topological luminophor $Y_2O_3$: $Eu^{3+}$+Ag was acquired by placing the $Y_2O_3$:$Eu^{3+}$+AgCl phosphors under sunlight for 72 h until the samples' color changed to gray with the decomposition of AgCl to Ag.

**2.4 Characterizations**

The X-ray diffraction (XRD) was performed on a Hitachi XRD-7000 diffractometer with Cu Kα irradiation. High-resolution transmission electron microscopy (HRTEM) images were obtained by a FEI Talos F200X transmission electron microscope equipped with an energy dispersive spectrometer (EDS). The surface size of a single sample was recorded using Bruker Dimension Fast Scan and Dimension Icon atomic force microscope (AFM). The electroluminescence (EL) structure is a typical sandwich structure and EL emission spectrum was measured through an Ocean Optics USB2000 fiber optic spectrometer.

**3 Results and discussion**

The phase composition of the final products was first examined by XRD. Fig. 3(a) shows the XRD pattern of topological luminophors $Y_2O_3$:$Eu^{3+}$+Ag prepared with different amounts of Ag. The XRD peaks of $Y_2O_3$:$Eu^{3+}$+Ag 0% were completely indexed to the cubic yttrium oxide $Y_2O_3$(PDF#41-1105) indicating that the $Eu^{3+}$ ions were doped into the $Y_2O_3$ structure without forming any detectable impurity phases. Other similar peaks indicated that the doped Ag did not change the crystal structure of $Y_2O_3$. The slight blue shift was caused by the Ag, which was doped in $Y_2O_3$, occupying a certain space at a great lattice constant. By contrast, the new peaks were indexed to the crystal silver Ag(PDF#41-1402). As shown in Fig. 3(b), the diffraction peaks of $Y_2O_3$:$Eu^{3+}$+AgCl 5% were different from those of $Y_2O_3$:$Eu^{3+}$+Ag 5% because of the AgCl(PDF#31-1238) inside $Y_2O_3$:$Eu^{3+}$ without any light treatment. The results indicate that most of the AgCl in the samples is decomposed into Ag through illumination and this feature was important for the asymmetric-discrete topological structure on the $Y_2O_3$:$Eu^{3+}$+Ag surface.

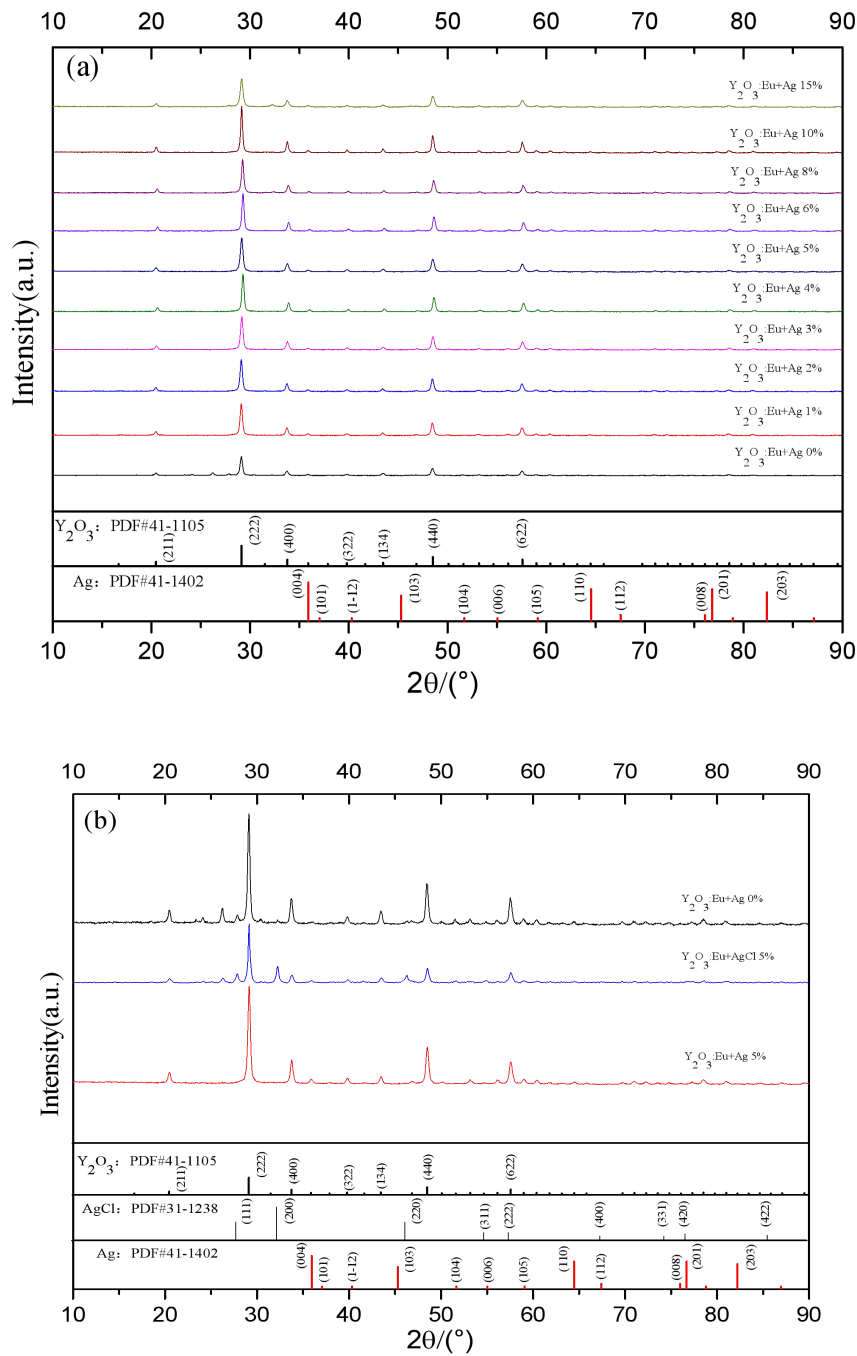

**Fig.3** XRD patterns of samples: (a) XRD spectra of topological luminophors $Y_2O_3$: $Eu^{3+}$+Ag with different contents of Ag. (b) XRD spectra of $Y_2O_3$: $Eu^{3+}$+AgCl 5%, $Y_2O_3$: $Eu^{3+}$+Ag 5% and $Y_2O_3$: $Eu^{3+}$ samples.

To assess the stability and area uniformity of a single sample, atomic force microscopy (AFM) imaging was performed at multiple locations across the samples (Fig. 4(a) and Fig. 4(b)). The AFM images showed a nomal sheet over a single sample that shared similarities to the HRTEM images and the size of topological luminophor

$Y_2O_3$:$Eu^{3+}$+Ag was not affected by illumination almostly. In particular, the average size of both $Y_2O_3$:$Eu^{3+}$+AgCl 5% and $Y_2O_3$:$Eu^{3+}$+Ag 5% was about 400 nm×300 nm×40 nm (Fig. 4(c)). The compositional evidence of the topological structure was further revealed by elemental mapping in (Fig. 4(d)). By contrast, the Ag color in a single microsheet indicated that AgCl was separated to Ag on the surface of topological luminophor $Y_2O_3$:$Eu^{3+}$+Ag asymmetrically and discretely and the element distribution was far different from Y or Eu. The Ag topological structure could be nearly regarded as a coated layer in terms of Chern count, and it was deeply connected to the luminescent performance of topological luminophor $Y_2O_3$:$Eu^{3+}$+Ag that based on the effect of surface plasma.

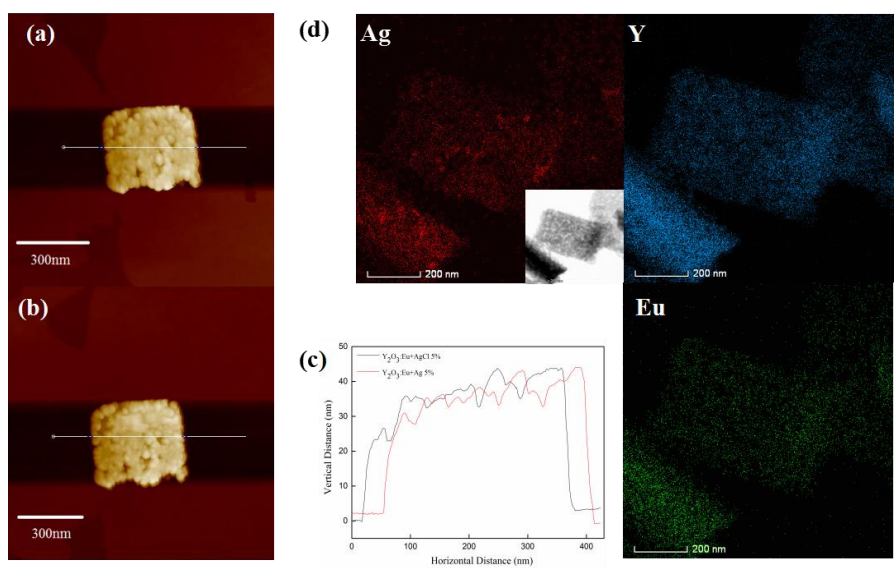

**Fig.4** (a) AFM image of single microsheet of $Y_2O_3$: $Eu^{3+}$+AgCl 5% sample. (b) AFM image of single microsheet of topological luminophor $Y_2O_3$: $Eu^{3+}$+Ag 5%. (c) Average size of single microsheet of both $Y_2O_3$: $Eu^{3+}$+AgCl 5% and $Y_2O_3$: $Eu^{3+}$+Ag 5% sample was about 400 nm×300 nm×40 nm. (d) Elemental mapping conducted on the single microsheet of topological luminophor $Y_2O_3$: $Eu^{3+}$+Ag 5% by energy dispersive spectrometer (EDS) including: Ag, Y and Eu.

The morphologies of samples were investigated by HRTEM imaging (Fig. 5). The morphology and size of microsheets were relatively uniform and unchanged by doped Ag or illumination. The deep color in the images are caused by overlapped small pieces. The observation of Ag via HRTEM was difficult because of the tiny size and content.

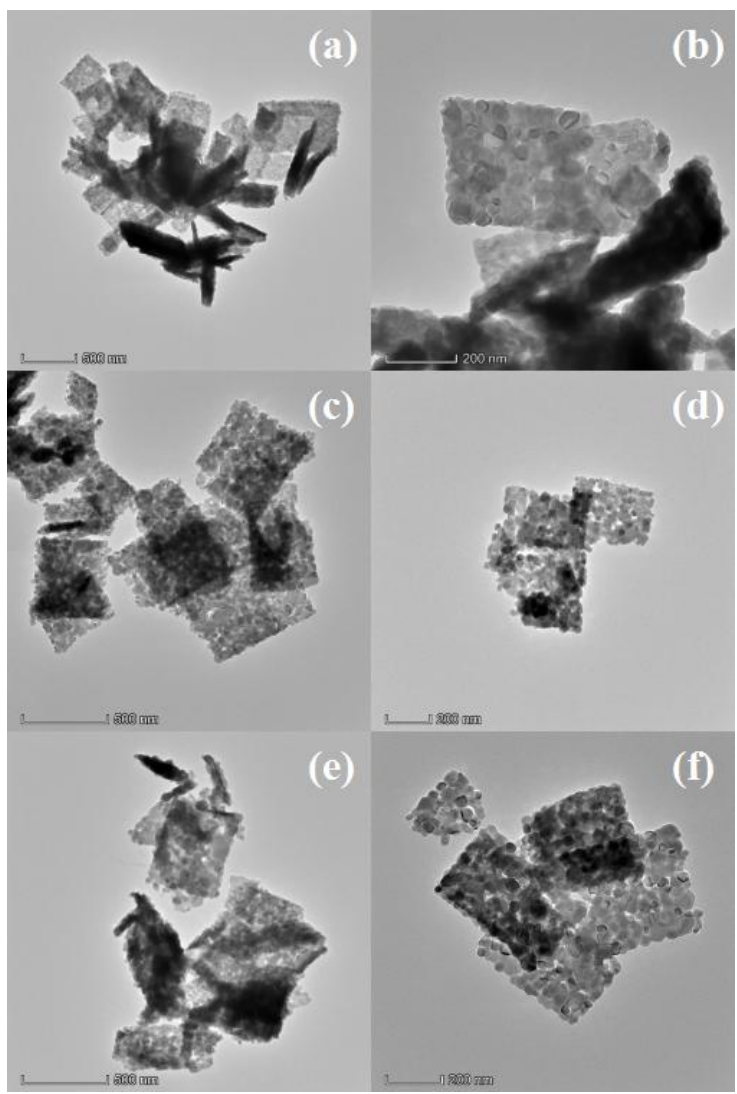

**Fig.5** HRTEM images of samples: (a), (b) $Y_2O_3:Eu^{3+}$ sample at different magnifications. (c), (d) $Y_2O_3:Eu^{3+}$+AgCl 5% sample at different magnifications. (e), (f) $Y_2O_3:Eu^{3+}$+Ag 5%  sample at different magnifications.

  The luminescent property under various electric field excitations of samples was investigated in detail. Fig. 6 shows that all the topological luminophors $Y_2O_3:Eu^{3+}$+Ag with different proportions of Ag demonstrated the strongest peak at 613 nm which was the luminescent characteristic peak. As Fig. 6(a) shows, the intensity at 800 V decreased from 273.83 to 279.01 as the molar ratio of Ag is increased from 0% to 1%, but peaked to 1248.06 as the Ag dosage increased to 5%. However, the intensity decreased drastically with the content increase from 6% to 15% possibly due to concentration-quenching and the miximum was 163.28 when the Ag dosage was 15%. The luminescent law of sample at 800 V was similar when the electric field increased to 1000 V or 1200 V (Fig. 6(b)). The luminescent intensity improved in Fig. 6(c), the Ag content increased from 0% to 5% at 1400 V and the maximum was 3312.78 when the dosage was 5%. The samples began to break down at 1200 V, and the Ag content exceeded 6%. Simultaneously, luminescent quenching was disrupted at 1400 V. These results suggested that the luminescent intensity of topological luminophors $Y_2O_3:Eu^{3+}+Ag$ is enlarged with the increase in the electric field and Ag content within a certain range. The details of luminescent property under various electric fields with diifferent Ag contents were showen in Fig. 6(d). All these findings were in accordance with the abovementioned luminescent law.

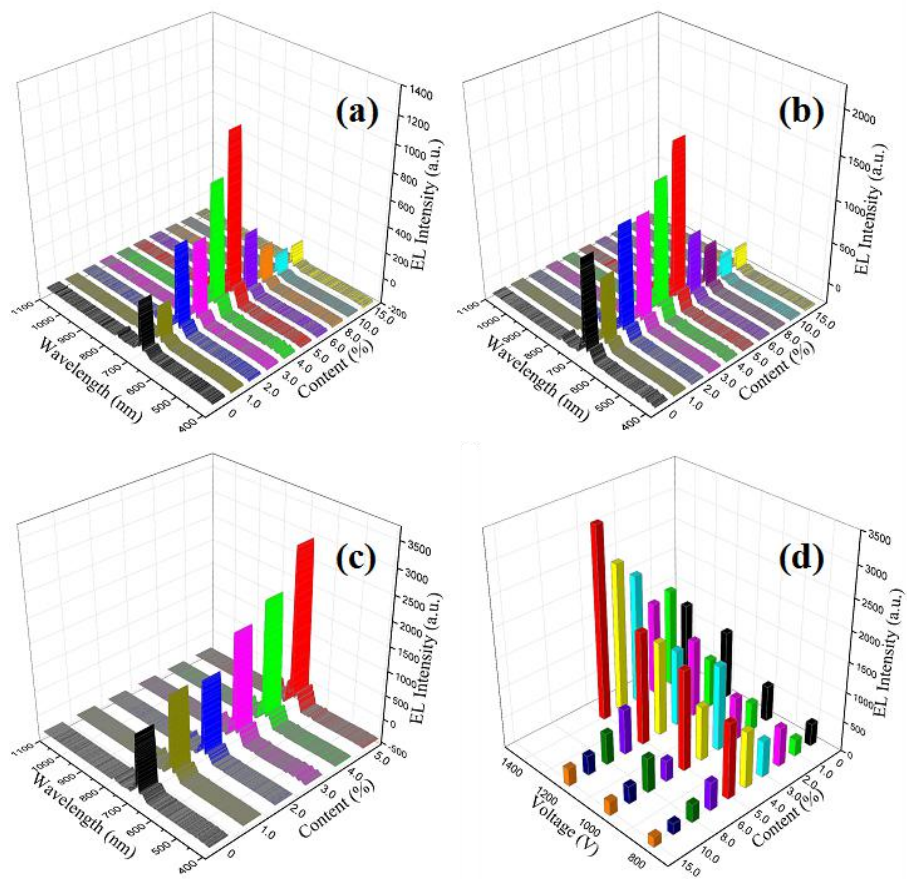

**Fig.6** EL spectra of topological luminophor $Y_2O_3:Eu^{3+}+Ag$ with different Ag contents at (a) 800 V, (b) 1200 V, (c) 1400 V.  (d) EL spectra for peak value of $Y_2O_3:Eu^{3+}+Ag$ topological luminophor with different Ag contents under various electric fields.

As shown in Fig. 7(a), stable luminescence of topological luminophor $Y_2O_3:Eu^{3+}+Ag$ 5% started from 800 V to 1400 V and the strongest peak was noted at 613 nm. The luminescent intensity strengtheded gradually with increasing of electric field and the maximum was obtained at 1400 V. In a stark contrast, luminescent intensity of $Y_2O_3:Eu^{3+}+AgCl$ 5% was the lowest among the three samples, but only slightly larger than the intensity of $Y_2O_3:Eu^{3+}+Ag$ 5%. This result was due to AgCl inside the microsheet, its inability to glow or the surface plasma resonance of metal nanoparticles (Fig. 7(b) and Fig. 7(c)). Illumination did not only change the surface structure of topological luminophor $Y_2O_3:Eu^{3+}+Ag$ but also substantially improved its luminescent intensity. The maximum luminescent intensity of topological luminophor $Y_2O_3:Eu^{3+}+Ag$ 5% was larger than that of $Y_2O_3:Eu^{3+}+AgCl$ 5% by two fold and three times than that of $Y_2O_3:Eu^{3+}+Ag$ 0%.

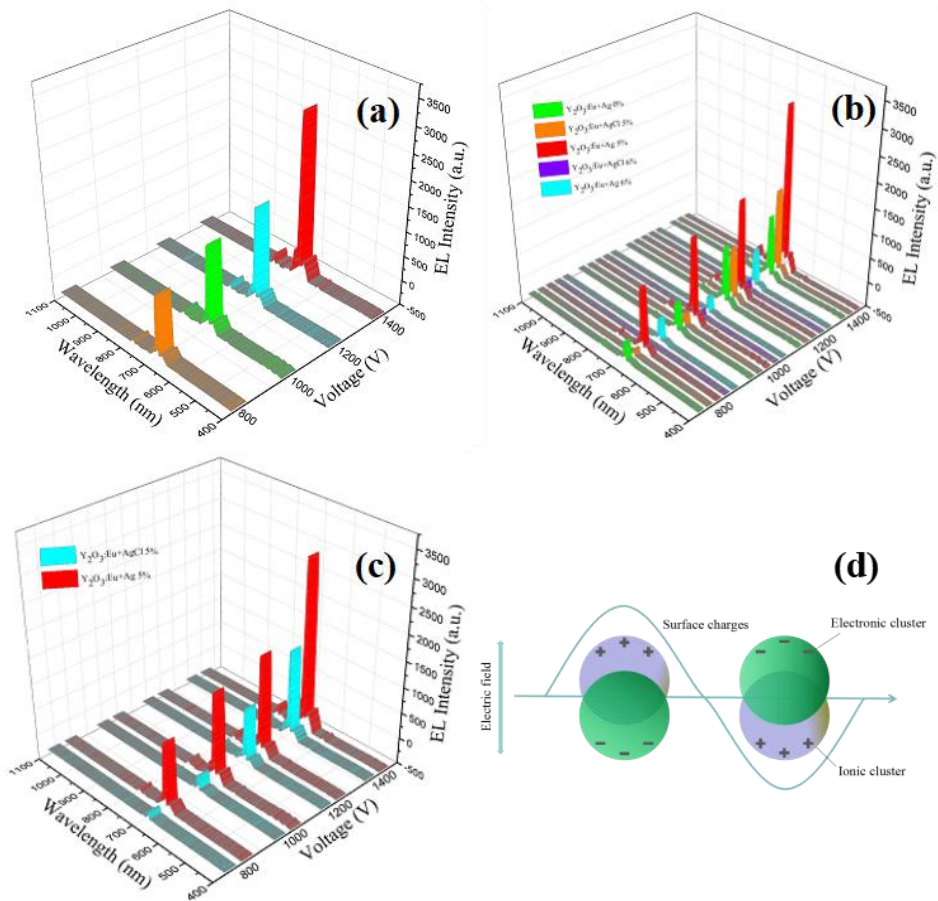

**Fig.7** (a) EL spectra of topological luminophor $Y_2O_3:Eu^{3+}+Ag$ under different electric fields. (b) Comparison of EL spectra before and after illumination at the Ag dosages of 0%, 5% and 6%. Note that the samples with illumination produced Ag nanocrystals with topological structure and high luminescent intensity, and the lluminescent intensity was proportional to the electric field for all the samples. Additionally, as the Ag content exceeded the best dosage(5%), the electroluminescent intensity reduced swiftly and the samples broke down under high electric field at the same time. (c) Comparison of EL spectra between $Y_2O_3:Eu^{3+}+AgCl$ 5% and $Y_2O_3:Eu^{3+}+Ag$ 5% samples. (d) Schematic representation of Ag nanocrystal topological structure induced by surface plasmons: $Eu^{3+}$ was treated as the luminescence center to produce electroluminescence under electric field which induced the Ag nanocrystal topological structure to generate surface plasmons.

A fundamental principle for the enhanced luminescent intensity of topological luminophor $Y_2O_3:Eu^{3+}+Ag$ could be generalized. In particular, the luminescence observed was not simple electroluminescence. $Eu^{3+}$ functioned as a luminescence center and was excited to produce electroluminescence under an electric field in the first place. $Eu^{3+}$ induced the Ag nanocrystal topological structure on the surface of $Y_2O_3:Eu^{3+}+Ag$ topological luminophor, to enhance photoluminescence of the samples based on surface plasma resonance (Fig. 7(d)). Eventually, it should be noted that the luminescence performance of topological luminophor $Y_2O_3:Eu^{3+}+Ag$ is increased by composite-luminescence which consisted of electroluminescence and photoluminescence.

## 3 Conclusion

A fundamental model for topological luminophor $Y_2O_3$:$Eu^{3+}$+Ag, which could be excited by composite-luminescence with $Eu^{3+}$centric electroluminescence and Ag surface plasma-enhanced photoluminescence was designed. On the basis of the hydrothermal preparation of $Y_2O_3$:$Eu^{3+}$ microsheet phosphors, $Y_2O_3$:$Eu^{3+}$+AgCl was synthesized by hydrothermal preparation. The asymmetric-discrete Ag nanocrystal topological structure was developed on the surface of $Y_2O_3$:$Eu^{3+}$ microsheets just via illumination. Instead of coating or magnetron sputtering of metallic surface structure, we used the relatively simple preparation technology of topological luminophor $Y_2O_3$:$Eu^{3+}$+Ag. $Eu^{3+}$ was treated as the luminescence center to produce electroluminescence under an electric field including Ag nanocrystal topological structure to enhance photoluminescence of the samples comply for surface plasma resonance. Thus, the luminescence performance of topological luminophor $Y_2O_3$:$Eu^{3+}$+Ag increased by about 300% by composite-luminescence which consisted of electroluminescence and photoluminescence. The proposed design of topological luminophor provides a new approach to further improve the luminescent intensity of phosphors.

**Acknowledgments**

This work was supported by the National Natural Science Foundation of China (Grant Nos. 11674267, 51272215).

**Conflict of Interest**

The authors declare no conflict of interest.